\documentclass[12pt,a4paper]{article}
\usepackage{graphicx}
\usepackage{amsmath} 
\newcommand{\avg}[1]{\langle #1\rangle}
\newcommand{\histoAvg}[1]{ \overline{\{#1 \} } }

\begin{document}
\textwidth=135mm
 \textheight=200mm
\begin{center}
{\bfseries Energy momentum conservation effects on two-particle correlation functions}
\vskip 5mm
N. Bock$^{\dag,\ddag}$
\vskip 5mm
{\small {\it $^\dag$ The Ohio State University, Columbus Ohio 43210, USA, }} 
\\
\end{center}
\vskip 5mm
\centerline{\bf Abstract}
Two particle correlations are used to extract information about the characteristic size of the system
in proton-proton and heavy ion collisions. The size of the system can be extracted from the Bose-Einstein 
quantum mechanical effect for identical particles. However there are also long range correlations  
that shift the baseline of the correlation function from the expected flat behavior. A possible source
of these correlations is the conservation of energy and momentum, especially for small systems, where the energy
available for particle production is limited. A new technique, first used by the STAR collaboration, of quantifying
these long range correlations using energy-momentum conservation considerations is presented in this talk.
Using Monte Carlo simulations of proton-proton collisions at 900 GeV,  
it is shown that the baseline of the two particle correlation function can be described using this technique.

\vskip 10mm
\section{\label{sec:intro}Introduction}
Femtoscopy is a powerful technique that allows us to "look" directly into the interaction region in 
particle collisions. It was first used in proton-antiproton collisions by Goldhaber,Goldhaber, Lee and Pais in 1969 \cite{GGLP} but it was developed by Hanbury Brown and Twiss in 1956 to measure the angular size of stars\cite{HBT} and is therefore also known as HBT. With this technique the size and lifetime of the emission region can be measured to study
the shape and evolution of its components. The measured radii and their dependence on pair momentum and multiplicity 
provide information about the level of interaction of the created particles, and can be used in particular to determine
the freezeout volume in heavy ion and proton-proton collisions. 

The HBT effect arises from the Bose-Einstein quantum mechanical effect for identical particles. However, any given pair of particles can be correlated through other effects, like resonance decays, jets and mini-jets, or phase-space constrains from energy momentum conservation. In this paper the latter are studied,  and it is organized as follows: a brief introduction to the femtoscopic techniques is given in section \ref{femto}, the energy momentum correlations are explained in section \ref{EMCIC} and their application to correlation functions are shown in section \ref{Fit}.

\section{Femtoscopy}
\label{femto}
\begin{center}
\begin{figure}[h]
\begin{minipage}{14pc}
Pairs with small relative momentum  $Q$ will be observed with higher rates because identical bosonic particles are likely to be found in the same state.
This gives rise to a peak at low momentum difference, and its width is inversely proportional to the size of the 
emission region. In particular for a Gaussian source $\rho(r) \sim \exp^{- r^2 / 2R^2}$ one obtains:
\end{minipage}\hspace{3pc}%
\begin{minipage}{14pc}
\includegraphics[width=16pc, height= 10pc]{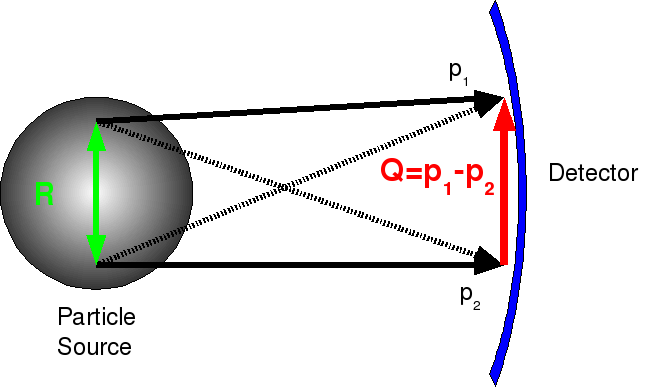}
\caption{\label{HBTFig}HBT effect.}
\end{minipage} 
\end{figure}
\end{center}

When dealing with two identical particles in quantum mechanics it is necessary to describe them with a 
symmetrized two-particle wavefuction:  
\begin{equation}
	\Psi(p_1,r_1,p_2,r_2)\sim e^{i (p_1\cdot r_1+ p_2 \cdot r_2)} +  e^{ i (p_1\cdot r_2+ p_2 \cdot r_1)}
\end{equation}
Wavefunction symmetrization leads to an enhanced probability of measuring pairs of particles with small momentum difference: 
\begin{equation}
	|\Psi^2|\sim  1+\cos((p_1-p_2)\cdot(r_1-r_2)) = 1+\cos(Q\cdot \Delta r),
\end{equation}
where $Q=p_1-p_2$ is the  momentum difference and $\Delta r = r_1-r_2$ is the relative emission point of the particles, 
see Fig.~\ref{HBTFig}. 
\begin{equation} 
\label{gaussian}
          C_{th}(Q)=1+\lambda e^{-R^2Q^2},
\end{equation}
where $\lambda$ is the correlation strength. The experimental correlation 
function is defined as the ratio of the two particle and single particle distributions:
\begin {equation}
	\label{CF}	
          C_{exp}(Q)=\frac{f(p_1,p_2)}{f(p_1)f(p_2)}
\end{equation}
Eqn.~\ref{gaussian} assumes the only correlation between particles is bosonic.
The ALICE collaboration measured Bose-Einstein correlations of pions at 900 GeV \cite{900GeVPaper}, and it was observed that the extracted radii depend on what type of baseline is used in the fitting procedure.
In this paper we are concerned about studying a possible source of non-femtoscopic correlations that can affect the measured radii, namely Energy-Momentum Conservation Induced Correlations (EMCICs). 

\section{Energy-momentum conservation induced correlations}
\label{EMCIC}
The phase-space available for particle production is constrained by energy momentum conservation, therefore 
creating a correlation among particles. It has been shown by Chajecki and Lisa \cite{Zibi}, that these correlations
can be quantified by applying energy-momentum constraints on the single particle distributions. This results
in an expression for the energy-momentum correlation of two particles:
\begin{equation}
\label{emcic1}
C_{EMCIC}(Q) = 1 - \frac{1}{N}\Bigg (2 \frac{ \mathbf{p_T}_{1} \cdot \mathbf{p_T}_{2}  }{\avg{p_T^2}}
              +\frac{p_{z_1} \cdot p_{z_2} } {\avg{p_z^2}} 
        	+\frac{(E_1-\avg{E})(E_2-\avg{E})   }{\avg{E^2}-\avg{E}^2}\Bigg)  
\end{equation}
where $N$ is the total event multiplicity. 
When femtoscopic correlations are present the total correlation function can be written as:
\begin{equation}
	\label{CFtotal}
   C(Q) = \Phi_{femto}(Q)\times C_{EMCIC}(Q)
\end{equation}
where $\Phi_{femto}$ is the femtoscopic or Bose-Einstein correlation. This expression implies that if all the quantities 
in Eqn.~\ref{emcic1} could be measured, the EMCICs would be completely characterized and it would be possible to remove their effect from the experimental correlation function. However, in experiment $N,\avg{p_T^2},\avg{p_z^2},\avg{E}$ and $\avg{E^2}$ cannot be measured, simply because not every particle is detected. Parametrizing the unknown quantities 
yields an equation that can be used to fit experimental data, as has been done in STAR for 200 GeV $p+p$ collisions  \cite{starPaper} and is being studied in ALICE for 900 GeV $p+p$ collisions \cite{HQBock}:

\begin{equation}
	\label{Cemcic}
  C(Q)= 1 - M_1\cdot\histoAvg{\mathbf{p_T}_{1}\cdot \mathbf{p_T}_{2}} - M_2 \cdot 
             \histoAvg{p_{z_1}\cdot p_{z_2}}-M_3\cdot\histoAvg{E_1\cdot E_2} + 
	M_4\cdot\histoAvg{E_1+E_2}-\frac{M_4^2}{M_3}
\end{equation}
The notation $\histoAvg{X}$ represents histograms of two-particle quantities that can be measured in experiment and are binned at the same time as the numerator and denominator of the correlation function in Eqn~\ref{CF}.
The $M_i$ parameters are related to the quantities that are not directly measurable in the following way:
      \begin{align}	
          M_1&=\frac{2}{N\avg{p_T^2}} \qquad \qquad \qquad	
          M_2=\frac{1}{N\avg{p_z^2}}\\	
          M_3&=\frac{1}{N(\avg{E^2}-\avg{E}^2)} \qquad	
          M_4=\frac{\avg{E}}{N(\avg{E^2}-\avg{E}^2)}	
        \end{align}
After a fit has been performed on the data, the physical quantities can be calculated 
using an additional equation for the energy of a characteristic particle with mass $m_*$ in the system :

\begin{equation}
  \avg{E^2}=\avg{p_T^2}+\avg{p_z^2}+m_*^2
\end{equation}
The total multiplicity of the event is then calculated as:
\begin{equation}
    N = \frac{M_3^{-1}-2M_1^{-1}-M_2^{-1}}{m_*^2-(M_4/M_3)^2}
\end{equation}
The physical quantities can be used as a consistency check to tell
if the fit results are reasonable.

\section{Baseline study of two-pion correlation functions.}
\label{Fit}
The shape of the baseline is studied using Monte Carlo simulations
of proton-proton collisions at 900 GeV, with a data set consisting of 10M events. 
The data analysis used to obtain all the histograms mentioned above, is done using the 
AliFemto code. A transverse momentum cut is applied to ensure correct 
identification of pions in a detector like ALICE, i.e. 0.1 GeV $< p_T <$ 1.2 GeV. 
The analysis also includes binning in four pair momentum bins $k_T = 1/2 |\mathbf{p_T}_{1}+\mathbf{p_T}_{2}|$, 
whose edges are 0.1,0.25,0.4,0.55,1.0 GeV.

\begin{center}
\begin{figure}[]
\centering
\includegraphics[width=16pc, height= 10pc]{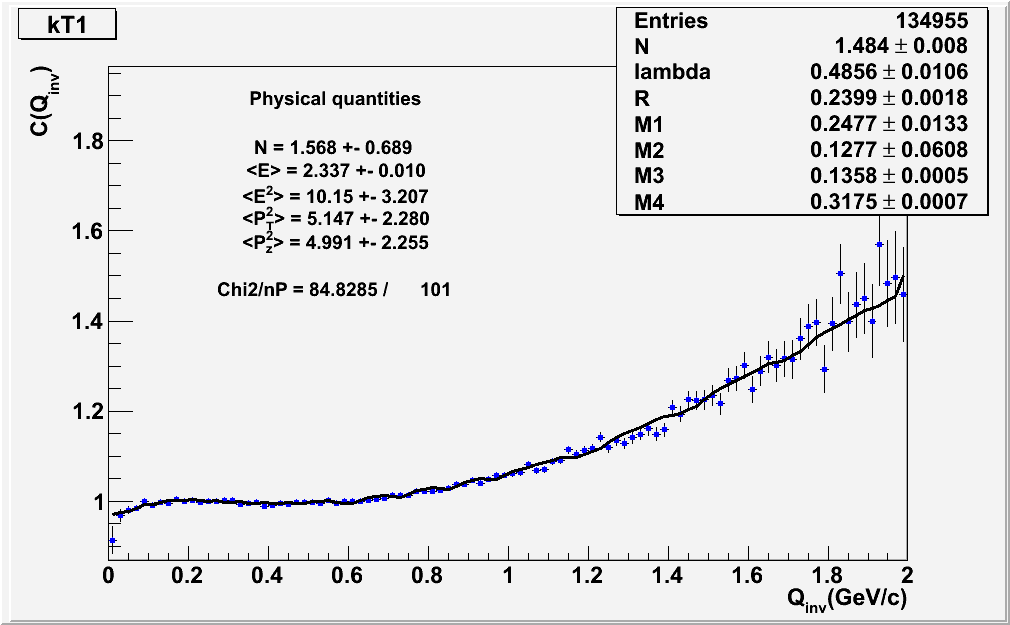}
\caption{\label{Qinv} EMCIC fit to the $Q_{inv}$ correlation function.}
\end{figure}
\end{center}

\begin{center}
\begin{figure}[]
\centering
\includegraphics[width=26pc, height= 18pc]{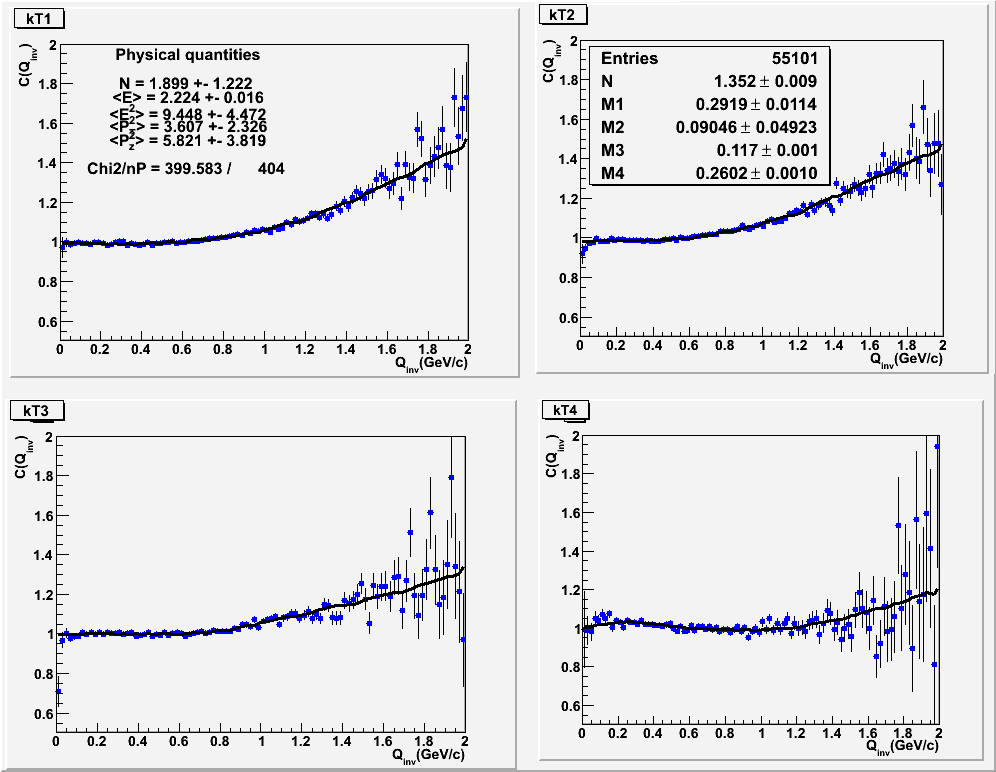}
\caption{\label{ktFit} EMCIC fit to the $Q_{inv}$ correlation function simultaneously in 4 $k_T$ bins.}
\end{figure}
\end{center}

A fit to the $Q_{inv}$ correlation function with no $k_T$ binning can be seen in Fig.~\ref{Qinv}.
The behavior at low $Q_{inv}$ is from two track effects, known as merging and splitting. At large $Q_{inv}$
the long rage correlation is very pronounced. The fitting function used is Eqn.~\ref{CFtotal}, including the Gaussian component to account for the underlying event at low $Q_{inv}$. The fit adjusts very well to the data, however,
it was observed that the $M_i$ values are not unique and depend on the initial values given to the parameters. 
This means that there are many local minima, which is expected from a one dimensional function with 7 parameters.
The physical quantities obtained from the $M_i$, also shown in Fig.~\ref{Qinv}, indicate that the fit is in a 
physical region. The average energy $<E> = 2.34 \pm 0.01 GeV$ is the most significant here given the small error bar, 
and it is a reasonable value. The total event multiplicity is much lower than expected and the average transverse momentum is quite high but both have a large error bar. 
A way to constrain the parameter space is to use the $k_T$ binning mentioned above and
fit all bins simultaneously with independent $N,\lambda,R_{inv}$ but the same $M_i$. This
is possible since the $M_i$'s are related to physical quantities which in this case are the
same for all $k_T$ bins. The results of this procedure are shown in Fig.~\ref{ktFit}.
The fitting function adjusts very good in all four $k_T$ bins, and it was observed that the convergence
of the $M_i$ values was more consistent when modifying initial values. As before
the average energy $<E> = 2.22 \pm 0.02 GeV$ is the most significant quantity, due to its small error bars, 
and it is again a reasonable value indicating a good fit. 
\section{Summary and future work}
A new technique to quantify and remove effects of energy-momentum conservation correlations from femtoscopic
effects has been used here to study the baseline of the $Q_{inv}$ correlation function. It was shown that the convergence
of fits to the data is dependent on initial parameters but it explains the long range correlations from first principles. 
Future work on this topic includes using the same procedure to fit correlation functions in the longitudinal co-moving system LCMS and in spherical harmonics, where there should be more constrains on the parameters allowing the fit
to converge easier.


\end{document}